\documentstyle[12pt,epsf]{article}

\newcommand{\rr}{\rightarrow}

\newcommand{\be}{\begin{equation}}
\newcommand{\ee}{\end{equation}}
\newcommand{\eel}[1]{\label{#1}\end{equation}}
\newcommand{\bea}{\begin{eqnarray}}
\newcommand{\eea}{\end{eqnarray}}
\newcommand{\eeal}[1]{\label{#1}\end{eqnarray}}
\newcommand{\baq}{\begin{equation}\begin{array}{rcl}}
\newcommand{\eaq}{\end{array}\end{equation}}
\newcommand{\eaql}[1]{\end{array}\label{#1}\end{equation}}
\newcommand{\beac}{\begin{equation}\begin{array}{rcl}}
\newcommand{\eeacn}[1]{\end{array}\label{#1}\end{equation}}
\newcommand{\ba}{\begin{array}}
\newcommand{\ea}{\end{array}}

\renewcommand{\d}{\delta}         
\newcommand{\e}{\varepsilon}

\newcommand{\beq}{\begin{eqnarray}}
\newcommand{\eeq}{\end{eqnarray}}

\newcommand{\bl}{\hspace{-.65cm}}
\newcommand{\gym}{g_{YM}}

\begin{document}
\newcommand{\preprint}[1]{\begin{table}[t]  
           \begin{flushright}               
           \begin{large}{#1}\end{large}     
           \end{flushright}                 
           \end{table}}                     
\preprint{hep-th/9904035}

\begin{center}
\Large{\bf A Comment on the Entropy of Strongly\\ Coupled ${\cal N}=4$}

\vspace{5mm}

\normalsize{N. Itzhaki}

\vspace{5mm}

{ Department of Physics\\
University of California, Santa Barbara, CA 93106}\\
{\it sunny@physics.ucsb.edu}

\end{center}


\begin{abstract}

We propose a field theory argument, which rests on the non-renormalization of
 the two point function of the energy-momentum tensor, why the ratio between the entropies of
 strongly coupled and weakly coupled ${\cal N}=4$ is of order one.

\end{abstract}


\baselineskip 18pt

The Maldacena conjecture \cite{mal} and the entropy of near-extremal D3-branes
\cite{gkp} imply that the ratio between the entropies, at fixed 
temperature, of strongly coupled and weakly coupled ${\cal N}=4$ is $3/4$.
In ${\cal N}=4$, unlike 2D CFT, the entropy is not protected thus it is not surprising that the
 ratio is not $1$.
It is surprising, however, that the ratio is not a function of the 't Hooft
coupling, $\lambda= \gym^2 N$,  which  vanishes when $\lambda \rr \infty$.

The reason is the following {\em perturbative} argument.\footnote{Though 
this argument is widely known we did not find
 it in the literature. A closely related  discussion can be  found  in 
\cite{hms,bjsv}.}
At finite temperature, $T$, the expectation value of the fields is 
$\langle \phi^2 \rangle =T^2.$
As a result the potential term in SYM, which has the form
$ V \sim \gym^2 [\phi_i, \phi_j ]^2,$
 induces a mass, $m^2 \sim \lambda T^2$, for a generic field.
At small 't Hooft coupling the induced masses are much smaller then the 
temperature so to a good approximation the contribution to the entropy is of 
$N^2$ massless fields with a small correction
which reduces the entropy.\footnote{For a  rigorous discussion 
 on the weakly coupled  region see \cite{ft}.}
At large 't Hooft coupling the induced masses are much larger then the
temperature.
Therefore, the contribution to the entropy from a generic field (not
in the Cartan subalgebra of $SU(N)$) is suppressed at the strongly coupled 
region.

Since the  argument above rests on perturbation theory  it cannot be
 trusted all the way to the strongly coupled region and hence, strictly 
speaking, there is no  contradiction with the  Maldacena conjecture.
Still, it is fair to say that it is somewhat disturbing that the only
field theory argument available (as far as we know) leads to that conclusion.
Especially, when a similar argument for SYM in $1+1$ dimensions 
\cite{hms,bjsv} leads to results that fit so nicely into the Maldacena
conjecture for D1-branes \cite{juan,pp,aki}.

The purpose of this short note is to put forward a {\em field theory}
argument, which does not rest on the AdS/CFT correspondence, 
that implies that the entropy at large coupling is of the order of the 
entropy at weak coupling. 
The argument rests on the ${\cal N}=4$ non-renormalization theorem for the
energy-momentum tensor two point function and therefore it cannot be generalised 
to two dimensional  SYM  which is a non-conformal theory  and hence the R-symmetry 
cannot protect the two-point function.

We study  SYM  in a box whose volume is 
$L_z A$ with $A=L_x L_y$ and we take the limit $L_x, L_y \gg L_z$.
Consider the transformation $x_3 \rr x_3(1-\e)$  with $\e \ll 1$.
The variation of the action under this transformation is 
\be
\d S=\e\int d^3 x \int_0^{L_z} dx_3 T_{33}.
\ee
Therefore, the variation of the expectation value of $T_{00}$ is
\beq
\d \langle T_{00}(0)\rangle =\int {\cal D} \phi ( e^{-(S+\d S)} -e^{-S}) 
T_{00}(0)=\e\int d^3x
\int_{0}^{L_z}dx_3
\langle T_{00}(0) T_{33}(x)\rangle,
\eeq 
where ${\cal D} \phi$ represents integration with respect to all fields.
To calculate the integral we need to know the energy-momentum tensor two 
point function.
On $R^4$ non-renormalization theorem protects
the energy-momentum tensor two point function. 
Thus on $R^4$ we can use the free SYM result
\be\label{11}
\langle T(0)T(x) \rangle =\frac{N^2}{x^8},
\ee
where  we have suppressed numerical factors of order one and the Lorentz 
indices (for details see \cite{2p}). 

However, what we  need is not the two-point function in $R^4$ but 
rather in  $R^3 \times S^1$.
In two dimensions the conformal transformation group contains the 
transformation from $R^2$ to $R \times S^1$.
Therefore, the two-points function in $R \times S^1$ are determined by  the 
two-points  function in $R^2$ and the dimensions of the operators. 
This is an important ingredient in Cardy's proof that the asymptotic
growth of the number of state of a 2D CFT depends only on the central charge
and not on the details of the CFT \cite{car,car1}.  
In four dimensions, however, the conformal transformations do not 
 contain the transformation from $R^4$ to $R^3 \times S^1$.
Thus, we do not know the exact form of the energy-momentum two point function
for strongly coupled SYM on $R^3 \times S^1$.\footnote{For weakly
 coupled theories  one can find directly on $R^3\times S^1$
the mode expansion of the relevant fields.
So there is no need to start with the two-points function on $R^4$.}

What we do know is that at distances smaller then $L_z$ 
the boundary condition is irrelevant and so
eq.(\ref{11}) is a good approximation at short distances.
Therefore, for a given point on $S^1$ we can calculate the contribution to
$\d \langle T_{00}(x_3)\rangle$ from the region $|x_3-x_3^{'}| < L_z$.
The integral divergent at short distances. The regularized result is 
$ \d \langle T_{00}(0)\rangle \sim \frac{\e N^2}{ L_z^4}$.
Integrating over the volume we find the variation of the ground 
state energy, which yields after 
integration with respect to $\d L_z=\e L_z$,  the ground state energy
\be\label{12}
E_0\sim \frac{N^2 A}{L_z^3}.
\ee
It is important to emphasis  that we have assumed in the calculation of 
$\d  \langle T_{00}\rangle$ that the integration
 over the whole region does not contain cancellations between the region where 
eq.(\ref{11}) is a good approximation 
and the region where it is not. 
Such cancellations can, in principle, reduce the ground state energy in a 
significant way to yield $E_0$ which is suppressed at large coupling.
Therefore, our argument is not a proof but rather a strong indication 
that the entropies ratio is of order one.
In other words, we estimate the Casimir energy, which is a boundary
condition effect, using an approximation which is not sensitive to the details
of the boundary condition but only to the  distance between the
 boundaries.

Eq.(\ref{12}) implies that the partition function at low temperature 
(compared to $L_z$) is,
\be
Z\sim \exp \left( \frac{N^2 A \beta}{L_z^3}\right) .
\ee
Now we can use the standard argument of switching the roles of 
$\beta $ and $L_z$ \footnote{See \cite{hm} for a related discussion in the context of 
the AdS/CFT correspondence.}  to end up with the partition function 
 of strongly coupled SYM at high temperature (compared to the size of the box)
\be
Z\sim \exp (N^2 V T^3),
\ee
which agrees, up to a numerical factor in the exponent, with the 
partition function of weakly coupled SYM.

\vspace{.5cm}

\bl {\bf Acknowledgements}

I would like to thank Aki Hashimoto for helpful discussions.
Work supported in part by the NSF grant PHY97-22022.

\end{document}